\documentclass[prd,aps,nofootinbib,preprintnumbers]{revtex4}

\usepackage{graphicx}
\usepackage[usenames]{color}
\usepackage{times}

\newcommand{\beq}{\begin{equation}}
\newcommand{\eeq}{\end{equation}}
\newcommand{\bea}{\begin{eqnarray}}
\newcommand{\eea}{\end{eqnarray}}
\newcommand{\bear}{\begin{array}}
\newcommand {\eear}{\end{array}}
\newcommand{\bef}{\begin{figure}}
\newcommand {\eef}{\end{figure}}
\newcommand{\bec}{\begin{center}}
\newcommand {\eec}{\end{center}}

\def\GEV#1{10^{#1}{\rm\,GeV}}

\begin{document}

\title{Flat Direction Inflation with Running Kinetic Term and Baryogenesis}

\author{Shinta Kasuya$^a$ and Fuminobu Takahashi$^{b,c}$}

\affiliation{
$^a$ Department of Mathematics and Physics,
     Kanagawa University, Kanagawa 259-1293, Japan\\
$^b$ Department of Physics, Tohoku University, Sendai 980-8578, Japan\\
$^c$ Kavli Institute for the Physics and Mathematics of the Universe (WPI), 
  Todai Institutes for Advanced Study, the University of Tokyo, Kashiwa 277-8582, Japan}

\date{May 16, 2014}

\begin{abstract}
We consider a possibility that one of the flat directions in the minimal supersymmetric standard model 
plays the role of the inflaton field and realizes large-field inflation.
This is achieved by introducing a generalized shift symmetry on the flat direction, 
which enables us to control  the inflaton potential over large field values. After inflation, higher order terms 
allowed by the generalized shift symmetry  automatically cause a helical motion of  the field  to create 
the baryon number of the universe, while baryonic isocurvature fluctuations are suppressed. 
\end{abstract}

\preprint{TU-970,~~IPMU14-0120}
\pacs{aaa}

\maketitle

%%%%%%%%%%%%%%%%%%%%%%%%%%%%%%%%%%%%%%
\section{Introduction}
Inflation solves various theoretical difficulties of the standard big bang 
cosmology~\cite{Guth:1980zm}.\footnote{
The exponentially expanding universe was also studied in Refs.~\cite{Brout:1977ix,
Starobinsky:1980te,Kazanas:1980tx,Sato:1980yn}.}
 The most plausible way to realize inflation with a graceful exit is to introduce 
a gauge singlet inflaton field with a relatively flat potential~\cite{Linde:1981mu,Albrecht:1982wi}.
 Indeed, the single-field slow-roll inflation
is consistent with observations including the cosmic microwave background (CMB) 
and large-scale structure data~\cite{Ade:2013uln}.

The BICEP2 collaboration recently reported detection of the primordial B-mode polarization
of CMB \cite{BICEP2}, which could be originated from
gravitational waves generated during inflation~\cite{Starobinsky:1979ty}.
The  tensor-to-scalar ratio  and the Hubble parameter during inflation suggested by the BICEP2
results are given by $r = 0.20^{+0.07}_{-0.05}$ and $H_{\rm inf} \simeq   \GEV{14} (r/0.16)^{1/2}$,
respectively.

Taken at face value, the BICEP2 results strongly suggest large-field inflation such 
as chaotic inflation \cite{Linde}, where the inflaton field excursion exceeds the reduced Planck scale, 
$M_{\rm P} \simeq 2.4 \times \GEV{18}$ \cite{Lyth:1996im}.  One way to control
the inflaton potential over such a broad field range is to introduce a
shift symmetry, under which the inflaton $\phi$ transforms as
\bea
\phi &\to& \phi +C,
\eea
where $\phi$ is the inflaton and $C$ is a real transformation parameter. Here and in what follows we adopt the Planck units where $M_{\rm P}$ is set to be unity unless explicitly shown for convenience.
The natural inflation~\cite{Freese:1990ni} or the multi-natural inflation~\cite{Czerny:2014wza,Czerny:2014xja}
is realized if the shift symmetry is explicitly broken by sinusoidal functions.
A similar shift symmetry was also used in the chaotic inflation in supergravity~\cite{Kawasaki:2000yn}.

Recently, a generalized shift symmetry was proposed~\cite{Linear,NT}:
\begin{equation}
\label{gss}
\phi^n \rightarrow \phi^n + C,
\end{equation}
where $n$ is a positive integer. In this case it is $\phi^n$ instead of $\phi$ that plays 
the role of the inflaton. Interestingly, this allows $\phi$ to be charged under gauge symmetry, because 
the inflaton remains gauge-singlet as long as $\phi^n$ is gauge invariant. As we shall see shortly, the 
generalized shift symmetry naturally leads to a field-dependent kinetic term, and as a result, the inflaton 
potential form changes at large field values. 

With the running kinetic term, the standard model (SM) Higgs field (as well as its supersymmetric
extension) can  play the role of the inflaton with a simple quadratic (or fractional power) 
potential~\cite{Nakayama:2010sk}. Similarly, one of the flat directions in the minimal supersymmetric 
standard model (MSSM) could play the role of the inflaton field and various implications
for phenomenology and cosmology were discussed in Ref.~\cite{NT}.\footnote{
See e.g. Refs.~\cite{Kasuya:2003iv,Allahverdi:2006we} for other realization of the MSSM flat direction inflation.}

In this Letter we revisit a possibility that one of the MSSM flat directions plays the role of the inflaton field. 
To this end we identify $\phi^n$ with the flat direction and impose the generalized shift symmetry (\ref{gss}).
Intriguingly, as pointed out in Ref.~\cite{Nakayama:2010sk}, 
the baryon or lepton-number violating operator required by the generalized shift symmetry 
necessarily induces a helical motion of the flat direction, creating the baryon asymmetry of
the universe {\it \`a la} Affleck-Dine mechanism~\cite{AD}. Thus, baryogenesis is a built-in feature
of the flat direction inflation with a running kinetic term. 
The flat direction condensate subsequently transforms into $Q$ balls \cite{KS,EM,KK}, 
and the reheating proceeds through the $Q$-ball decay \cite{Qsurf}. Since the flat direction consists of the supersymmetric partner of quarks and/or leptons, the reheating of the SM degrees of freedom is straightforward.
It is noteworthy that 
the baryonic isocurvature perturbations are suppressed in our inflation model, in contrast to 
the usual case \cite{KKT}. Thus we investigate the flat direction inflation with a running kinetic term in 
great detail, focusing on the baryogenesis due to the flat direction.

The structure of the Letter is as follows. In the next section, we review the running kinetic inflation in general.
We see how the flat direction fits into the context of the running kinetic inflation in Sec.III, where
the dynamics of the field is also investigated numerically. In Sec. IV, the baryon number of the 
universe is estimated, and we conclude in Sec.~V. In Appendix, we show a list of possible 
flat directions together with the predicted values of the scalar spectral index and the tensor-to-scalar 
ratio.

%%%%%%%%%%%%%%%%%%%%%%%%%%%%%%%%%%%%%%
\section{Running kinetic inflation}
Let us briefly review the running kinetic inflation in supergravity \cite{Linear,NT}. 
We introduce a chiral superfield $\phi$, whose K\"ahler potential is invariant under the generalized
shifts symmetry (\ref{gss}). The K\"ahler potential thus
is a function of $(\phi^n-\phi^{\dagger n})$, and we expand it as
\begin{equation}
K = \sum_{k =1} \frac{c_k}{k!} (\phi^n-\phi^{\dagger n})^k
= c_1(\phi^n-\phi^{\dagger n})-\frac{1}{2} (\phi^n-\phi^{\dagger n})^2 + \cdots,
\label{Kahler1}
\end{equation}
where $c_k$ is a numerical coefficient of $O(1)$.
In particular $c_1$ is a pure imaginary number and we normalize $c_2\equiv -1$. 
During inflation, the field stays along the inflationary trajectory which minimizes 
the K\"ahler potential (\ref{Kahler1}). For simplicity we neglect higher order terms with $k \ge 3$,
as they do not change the following argument.

For successful inflation, we introduce shift-symmetry breaking terms in the superpotential and K\"ahler potential as
\begin{eqnarray}
& & W = \lambda X \phi^m, \\
& & \Delta K = \kappa |\phi|^2,
\end{eqnarray}
where $\lambda$, $\kappa \ll 1$. As we shall see below,  $\lambda$ is fixed by the  normalization 
of curvature perturbations. We have introduced $X$ for successful inflation at large field 
values \cite{Kawasaki:2000yn}, and it can be stabilized at the origin with a positive Hubble-induced mass 
during inflation by higher-order terms in the K\"ahler potential.

The total superpotential and K\"ahler potential are given respectively by\footnote{
Note that, since the shift symmetry is only  approximate, there could be (infinitely) many shift-symmetry
breaking terms that are consistent with gauge (or other) symmetries.  If the shift symmetry is of high quality, 
however, the higher order terms can be suppressed by powers of the order parameters such as $\kappa$ or 
 $\lambda$. For instance,  the quantum gravity corrections to the inflaton potential are suppressed by 
certain powers of the order parameters, because the gravity is necessarily coupled to the
energy momentum tensor, which is accompanied by the order parameters. It is in principle possible that
 both $X \phi^m$ and $X \phi^{2m}$ terms play an important role during the last $50-60$  e-foldings. 
 In this case, the inflaton dynamics will be similar to the so-called polynomial chaotic inflation~\cite{Nakayama:2013jka}, and various values of $n_s$ and $r$ can be realized.
}
\begin{eqnarray}
\hspace{-8mm} & & 
\label{kaehler}
K = \kappa |\phi|^2 + c_1(\phi^n-\phi^{\dagger n})-\frac{1}{2} (\phi^n-\phi^{\dagger n})^2 +|X|^2 +\cdots, \\
\hspace{-8mm} & & W = \lambda X \phi^m,
\label{superpot}
\end{eqnarray}
where the dots represent higher order terms. Then the effective Lagrangian relevant for the inflation becomes
\begin{equation}
{\cal L} = (\kappa + n^2 |\phi|^{2n-2})\partial_\mu \phi \partial^\mu \phi^\dagger - V,
\end{equation}
where $V$ is the supergravity potential given by
\begin{equation}
V = e^{\kappa|\phi|^2+c_1(\phi^n-\phi^{\dagger n}) -\frac{1}{2}(\phi^n-\phi^{\dagger n})^2} 
\lambda^2 |\phi|^{2m}.
\label{sugraV}
\end{equation}
Here we omit the soft supersymmetry (SUSY) breaking terms for the inflaton. 
Notice that $X$ is assumed to be stabilized at the origin, $X=0$, during inflation. 
For $|\phi| \gtrsim (\kappa/n^2)^{1/(2n-2)}$,  the $\kappa$-term is negligible, and therefore, 
it is $\hat{\phi} \equiv \phi^n$ that is the canonically normalized field. The real component of $\hat{\phi}$
can take a super-Planckian field value and therefore becomes the inflaton, as it does not appear 
in the K\"ahler potential because of the shift symmetry. On the other hand, the imaginary component 
of $\hat{\phi}$ acquires a mass of order Hubble parameter, and it is stabilized where the K\"ahler potential is 
minimized, $\hat{\phi}-\hat{\phi}^\dagger = c_1$ for ${\rm Re}[\hat{\phi}] \gtrsim 1$.
This equation determines the inflationary trajectory for ${\rm Re}[\hat{\phi}] \gtrsim 1$. The fact that the 
imaginary component has a mass of order Hubble parameter will be important to suppress the isocurvature 
perturbations, when applied to the MSSM flat direction. 
Along the inflationary trajectory, the effective Lagrangian is given by
\begin{equation}
{\cal L} = \frac{1}{2} (\partial \hat{\phi}_R )^2 - \hat{\lambda}^2 (\hat{\phi}_R)^{2m/n},
\end{equation}
where $\hat{\phi}_R \equiv {\rm Re}[\hat{\phi}]/\sqrt{2}$, and 
$\hat{\lambda}^2 \equiv (e^{-|c_1|^2/2}\lambda^2)/2^{m/n}$. This is nothing but the chaotic inflation with
a monomial potential $\hat{\phi}^p$ with $p=2m/n$, and one can express the coupling $\lambda$, 
the spectral index $n_s$  and the tensor-to-scalar ratio $r$ respectively as
\begin{eqnarray}
\label{lamval}
\label{lambda}
& & \lambda \simeq 5.1\times 10^{-4} \,
e^\frac{|c_1|^2}{4} 2^{-\frac{m}{2n}} 
\left(\frac{m}{n}\right)^{\frac{n-m}{2n}}
N^{-\frac{m+n}{2n}}, \\
& & n_s = 1 -\left( 1+\frac{m}{n}\right)\frac{1}{N}, \\
& & r = \frac{8m}{n}\frac{1}{N},
\label{t2s}
\end{eqnarray}
where $N$ is the e-folding number, and we have used  in Eq.(\ref{lamval}) the Planck normalization on 
the curvature perturbations, $\Delta^2_{\cal R} \simeq 2.2 \times 10^{-9}$~\cite{Ade:2013uln}.

Inflation ends when the canonically normalized inflaton $\hat{\phi}_R$ becomes comparable
to unity. Then  the inflaton field leaves the inflationary trajectory, and starts rotation because 
of the $\phi$-number violating terms in Eq.~(\ref{sugraV}). Note that such $\phi$-number violating
terms are allowed by the shift symmetry, and indeed,  one can see from Fig.~\ref{fig1}(b) that, 
as long as $c_1 \ne 0$, the inflationary trajectory is off-set along the imaginary component of $\hat{\phi}$, 
leading to the rotation after inflation. On the other hand, if $c_1 = 0$, the inflationary trajectory coincides
with the real axis, and no rotation is induced after inflation. Therefore we expect that the resultant 
$\phi$-number is an increasing function of $c_1$. In fact, we verify this numerically, although the 
$c_1$-dependence is nontrivial.

After the amplitude of the field decreases below $|\phi| < (\kappa/n^2)^{1/(2n-2)}$, $\phi$ itself becomes 
the dynamical variable, and it rotates in the potential $|\phi|^{2m}$ until the soft SUSY breaking mass 
term $m_\phi^2|\phi|^2$ dominates the potential. Finally the inflaton decays into radiation, 
but this process is highly model-dependent.

%%%%%%%%%%%%%%%%%%%%%%%%%%%%%%%%%%%%%%
\section{Flat direction as inflaton}
In MSSM, flat directions are composed of squarks and sleptons. The potential is flat 
in the SUSY limit at renormalizable level, but lifted by SUSY breaking effects and 
nonrenormalizable terms. The former leads to a soft mass
term, while the latter gives rise to higher order terms in the potential. 

In order to see how to fit the flat direction in the context of the running kinetic inflation, we first specify
the direction. Flat directions are represented by gauge-invariant (GI) monomials and
they are all classified in Ref.~\cite{GKM}. Therefore, the flat direction is a good candidate of the
inflaton $\hat{\phi}=\phi^n$, where $\hat{\phi}$ should be gauge-singlet. In the previous section
we have not included the gauge interactions of $\phi$. In fact, the gauge bosons coupled to the
flat direction acquire a heavy mass at large fields values of $\phi$. For the parameters of our interest,
however, the physical gauge boson masses as well as the field value of $\phi$ do not exceed the Planck scale,
even though the canonically normalized inflaton $\hat{\phi}$ does become super-Planckian. Also,
$\hat{\phi}$ has only suppressed interactions with the gauge fields and other SM fields because
of the running kinetic term. Therefore, the argument in the previous section can be applied to
the MSSM flat directions. 

For successful inflation in supergravity, as mentioned 
in the previous section, the superpotential must have the form of (\ref{superpot}).\footnote{%%
The other degrees of freedom $Q_i$  can be stabilized at the origin, 
if there are $|X|^2|Q_i|^2$ terms in the K\"ahler potential.}
This is the case 
for the $d$ directions in Ref.\cite{GKM}, where $d=m+1$. We list some of these directions and the 
nonrenormalizable superpotential that lifts the corresponding direction in Table \ref{flat_table}.
For these directions, the inflation takes place effectively for $V \sim \hat{\phi}^p$ with $p=1.6 - 3.2$.
A complete list of the flat directions and the predicted values of $n_s$ and $r$ are given in the Appendix.
Note that, although we do not necessarily impose the matter parity here, we can do so by
considering the generalized shift symmetry on the gauge monomial squared. For instance, in the case of $LLe$,
we can impose the generalized shift symmetry on $(LLe)^2$ instead of $LLe$ so that the matter parity
is kept even for $c_1 \ne 0$ in  Eq.~(\ref{kaehler}). In this case, the actual value of $n$ is obtained by multiplying
$n'$ in Table 1 or 2 by a factor of two. 

%%%%%%%%%%%%%%%%%%%%%%%%%%%%%%%%%%%%%%%%%
\begin{table}[h!]
\caption{Examples of the flat direction candidates. $n = n'$ or $2n'$.}
\begin{tabular}{|c|c|c|}
\hline
Direction & GI monomials & $W_{\rm NR}$ \\
\hline
$L$, $d$, $e$ & $\left\{\begin{array}{l} LLe \ (n'=3) \\ LLddd \ (n'=5) \end{array} \right.$ 
& $\left\{\begin{array}{l} H_d Lddd \ (m=4) \\ H_uLLLe \ (m=4) \end{array} \right.$  \\
\hline
$L$, $d$ & $LLddd$ ($n'=5$) & $H_u LLLddd$ ($m=6$) \\
\hline
$Q$, $u$, $e$ & $QuQue$ ($n'=5$) & $H_dQuQuQuee$ ($m=8$) \\
\hline
\end{tabular}
\label{flat_table}
\end{table}
%%%%%%%%%%%%%%%%%%%%%%%%%%%%%%%%%%%%%%%%%%

Now we show numerical results of the dynamics of the inflaton field for $n=3$ and $m=4$, for example. 
In this case, the inflationary trajectory is represented as
\begin{equation}
3\phi_1^2\phi_2-\phi_2^3=\frac{|c_1|}{2}=\frac{1}{2},
\label{traj}
\end{equation}
where $\phi=\phi_1 + i\phi_2$ ($\phi_1$, $\phi_2$ is real), and we take $c_1 = i$ in the last equality.

\begin{figure}[ht!]
\begin{center}
\hspace*{11mm}
\vspace{10mm}
\begin{tabular}{ccc}
\includegraphics[width=80mm]{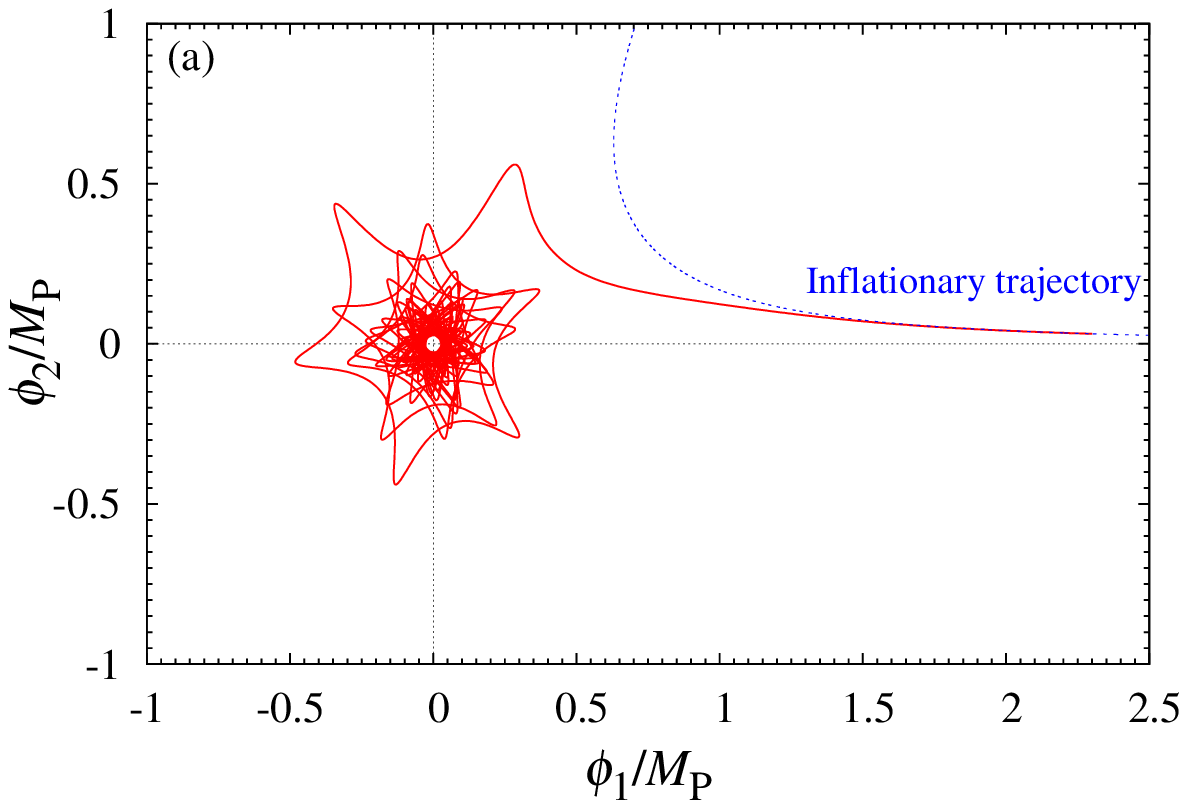} & \hspace{5mm} &
\includegraphics[width=80mm]{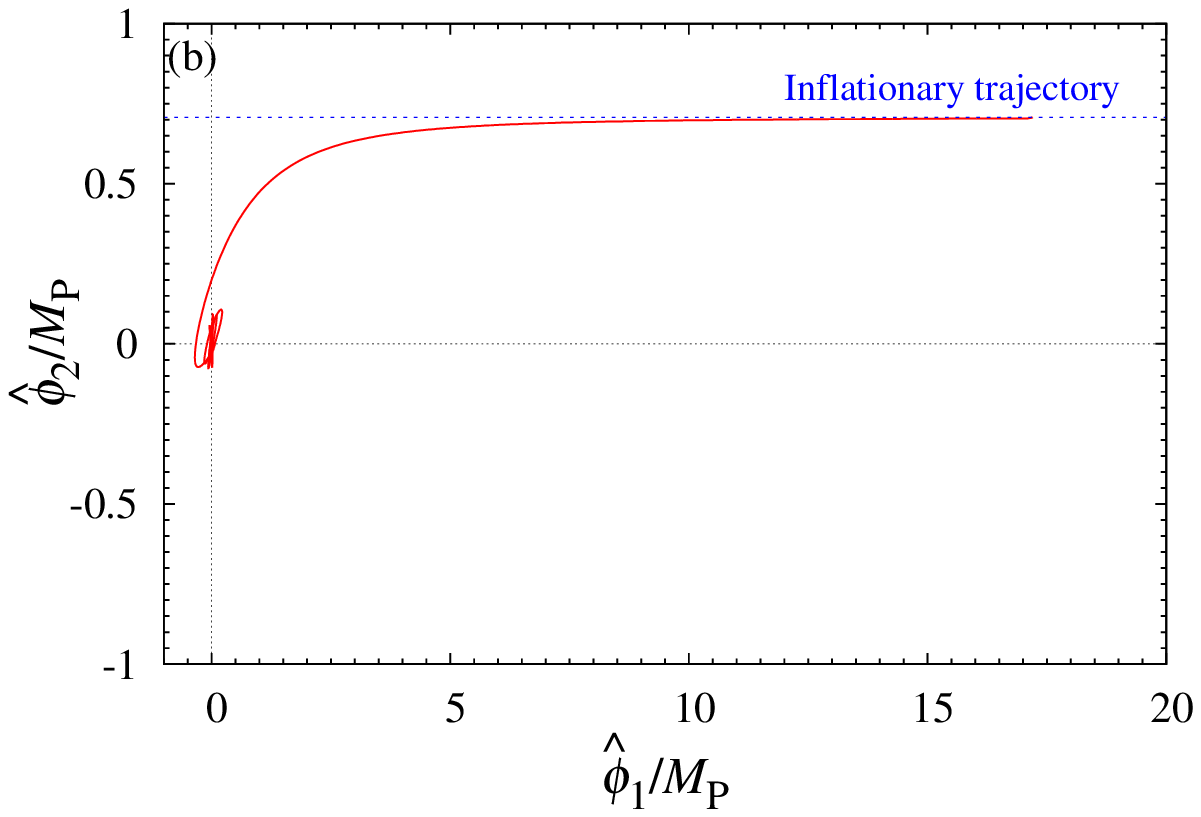} 
\end{tabular}
\caption{Left: Trajectory of the field $\phi$ for $n=3$ and $m=4$. 
We set $\lambda=2.8\times10^{-6}$ and $\kappa=0.01$. 
Right: Trajectory of the field $\hat{\phi}$ for the same dynamics.
\label{fig1}}
\end{center}
\end{figure}

In Fig.~\ref{fig1}(a), the trajectory of the field $\phi$ is shown. We set the initial conditions as
$\phi_1 = 2.3$ and $\phi_2=0.0315$, from where the inflation lasts for $N\simeq 58$ e-foldings.  
The field first evolves along the inflationary trajectory (\ref{traj}), and leaves from there when $\phi \sim 1$.
We also take the values of $\lambda=4.1 \times 10^{-6}$ (for $N=50$) and $\kappa=0.01$, which is consistent
with the Planck normalization on the curvature perturbations. In terms of the canonical normalized 
field for large amplitudes, the dynamics of the field $\hat{\phi} \, (=\phi^3)$
looks more familiar, shown in Fig.~\ref{fig1}(b).

%%%%%%%%%%%%%%%%%%%%%%%%%%%%%%%%%%%%%%%%%
%%%%%%%%%%%%%%%%%%%%%%%%%%%%%%%%%%%%%%%%%%

After inflation, the field starts rotation due to the $\phi$-number violating terms in K\"ahler potential. 
Drawing-star-like behavior will end when the field amplitude decreases to $\phi \sim \kappa^{1/(2n-2)}$.
Afterward, the canonically normalized field is represented by 
$\varphi$, where $\phi=\varphi e^{i\theta}/\sqrt{2\kappa}$, and the Lagrangian is given by
\begin{equation}
{\cal L}= \frac{1}{2}(\dot{\varphi}^2+\varphi^2\dot{\theta}^2) + \frac{1}{2} m_\varphi^2 \varphi^2
+\frac{\lambda^2}{(2\kappa)^m} \varphi^{2m},
\end{equation}
where the second term is the soft SUSY breaking mass term. The field will rotate in the potential 
of the form $V \sim \varphi^{2m}$, until the soft SUSY breaking mass term dominates the potential.

%%%%%%%%%%%%%%%%%%%%%%%%%%%%%%%%%%%%%%
\section{Baryogenesis}
The soft SUSY breaking mass term is actually given by
\begin{equation}
V_{\rm soft} = \frac{1}{2} m_\varphi^2 \varphi^2 \left( 1+ K \log \frac{\varphi^2}{2 M^2}\right), 
\end{equation}
where $K$ is a numerical coefficient of the one-loop radiative corrections and it becomes negative 
when the gaugino loop dominates \cite{EM}, and  $M$ is the renormalization scale where $m_\varphi$ 
is evaluated. In the gravity-mediation or anomaly mediation with a generic K\"ahler 
potential~\cite{Giudice:1998xp}, we expect that the soft mass is comparable to the gravitino mass, 
$m_\varphi \sim m_{3/2}$. We assume the gravity mediation in the following, but our results can be 
straightforwardly applied to the the pure gravity mediation scenario~\cite{Ibe:2011aa}, or the minimal 
split SUSY~\cite{ArkaniHamed:2012gw,Arvanitaki:2012ps}. The typical value of $K$ is  
$|K| = 0.01 - 0.1$ for the gravity mediation, while $|K|$ can be a few orders of magnitudes smaller 
if the gaugino masses are one-loop suppressed with respect to the sfermon masses~\cite{Higaki:2012ba}.

Once the amplitude of the field gets smaller than $\varphi_{\rm eq}$, where
\begin{equation}
 \varphi_{\rm eq} = \left(\frac{(2\kappa)^m}{2\lambda^2}\right)^{1/(2m-2)} m_\varphi^{1/(m-1)},
\end{equation}
the field experiences spatial instabilities due to the negative value of $K$ and transforms into 
$Q$ balls \cite{KS,KK}. The charge of the formed $Q$ ball is estimated as \cite{KK}
\begin{equation}
Q = \beta \left(\frac{\varphi_{\rm eq}}{m_\varphi}\right)^2,
\end{equation}
where $\beta=0.02$ \cite{Hiramatsu}.  

The universe becomes dominated by the $Q$ balls, and
the reheating is induced by the $Q$-ball decay. The $Q$ ball decays through its surface, 
and the rate has an upper bound called the saturated rate, which stems from the Pauli's blocking 
of the produced fermions \cite{Cohen}. The decay rate is estimated as \cite{KY,KKY}
\begin{equation}
\Gamma_Q = \frac{N_q}{Q} \frac{\omega_Q^3}{12\pi^2} 4\pi R_Q^2, 
\end{equation}
where $N_q$ is the number of the produced quarks by the decay. $\omega_Q$
is the effective mass of the $\varphi$ particle inside the $Q$ ball, and $R_Q$ is the size of the $Q$ ball.
They are given respectively by \cite{EM}
\begin{equation}
\omega_Q \simeq m_\varphi, \qquad R_Q \simeq |K|^{-1/2} m_\varphi^{-1}.
\end{equation}
Then the reheating temperature is obtained as
\begin{eqnarray}
\hspace{-7mm} & & 
T_R =  \left(\frac{90}{4\pi^2 g_*}\right)^{1/4} \sqrt{\Gamma_Q M_{\rm P}} \nonumber \\
\hspace{-7mm} & & 
\hspace{6mm} \simeq 370 \, {\rm GeV} \left(\frac{g_*}{96.25}\right)^{-1/4}
\left(\frac{m_\varphi}{10^4\, {\rm GeV}}\right)^{7/6} 
\left(\frac{N_q}{18}\right)^{1/2}
\left(\frac{|K|}{0.01}\right)^{-1/2} \left(\frac{\beta}{0.02}\right)^{-1/2}
\left(\frac{\lambda}{4.1 \times 10^{-6}}\right)^{1/3} \left(\frac{\kappa}{0.01}\right)^{-2/3},
\end{eqnarray}
where $g_*$ counts the relativistic degrees of freedom and we set $m=4$ in the last equality. 
Applying this reheating temperature, we can estimate the baryon number of the universe as 
\begin{eqnarray}
\hspace{-7mm} & &
Y_b  =  \frac{n_b}{s} = \frac{3}{4} T_R \left.\frac{n_b}{\rho_r}\right|_{\rm D}  
= \frac{3}{4} T_R \left.\frac{n_b}{\rho_\varphi}\right|_{\rm eq} \simeq \frac{3}{4} T_R \frac{1}{m_\varphi},
\nonumber \\ 
\hspace{-7mm} & & \hspace{6mm}
\simeq   2.8\times 10^{-2}  \left(\frac{g_*}{96.25}\right)^{-1/4}
\left(\frac{m_\varphi}{10^4\, {\rm GeV}}\right)^{1/6} 
\left(\frac{N_q}{18}\right)^{1/2}
\left(\frac{|K|}{0.01}\right)^{-1/2} \left(\frac{\beta}{0.02}\right)^{-1/2}
\left(\frac{\lambda}{4.1 \times 10^{-6}}\right)^{1/3} \left(\frac{\kappa}{0.01}\right)^{-2/3}.
\end{eqnarray}
Therefore, we can generate a sufficiently large amount of the baryon asymmetry from the
inflaton dynamics. Note that, since the imaginary component of $\hat{\phi}$ has a mass of order
Hubble parameter during inflation, there is no light degree of freedom other than the inflaton, 
which acquires quantum fluctuations during inflation. Therefore, there is no baryonic isocurvature
perturbations, in contrast to the usual Affleck-Dine mechanism~\cite{KKT}.

The baryon asymmetry estimated above is much larger than the observed value of $Y_b \sim 10^{-10}$,
but the resultant baryon asymmetry can be suppressed in either of two ways. As mentioned before,
the final baryon asymmetry can be suppressed if $|c_1|$ is sufficiently small. The smallness of $|c_1|$ can be
understood if $\phi^n$ has an odd matter parity, as it would represent the small breaking of the matter parity.
We have indeed confirmed numerically that the final baryon asymmetry can be suppressed for a sufficiently 
small value of $|c_1|$, although it is nontrivial to derive its analytic dependence because of the
complicated dynamics during oscillations (see Fig.~\ref{fig1}). On the other hand, there may
be late-time entropy production by a modulus decay of thermal inflation~\cite{Yamamoto:1985rd}, 
which dilutes the final baryon asymmetry.
For example, if the axion is the main component of the dark matter, the Peccei-Quinn symmetry should be 
restored during inflation to avoid the isocurvature bounds~\cite{Linde:1990yj,Lyth:1992tx}. In this case, thermal inflation may naturally takes
place by the saxion \cite{Choi:1996vz}.

Lastly let us comment that the reheating temperature is rather low thanks to the reheating through 
large $Q$ balls \cite{Qsurf}. If the reheating process had taken place through perturbative decay 
with the rate $\Gamma=\frac{f^2}{8\pi} m_\varphi$, where $f$ generally denotes a coupling constant, 
the reheating temperature would be estimated as
\begin{equation}
T_R = 7 \times 10^9 \, {\rm GeV} \left(\frac{g_*}{200}\right)^{-1/4}
\left(\frac{m_\varphi}{10^3\, {\rm GeV}}\right)^{1/6} 
\left(\frac{f}{0.1}\right).
\end{equation}
Similarly, if the decay proceeds through nonperturbative particle production such as preheating 
and/or thermal dissipation, the reheating temperature will be high.\footnote{%%
For perturbative or nonperturbative decay, the relativistic degrees of freedom $g_*$ might be 
altered due to the large effective mass of the particles coupled to the inflaton.}
If the reheating temperature is high,
the resultant baryon asymmetry tends to be large, which would require late-time entropy production. 
Note that the dependence of the final baryon asymmetry on $|c_1|$ could be involved, because, 
for small values of $c_1$, nonperturbative effects are expected to be more efficient, whereas the 
baryon-number violating operator is suppressed. In any case, it will be possible to generate
the right amount of the baryon asymmetry for sufficiently small values of $c_1$ since no baryon asymmetry
is generated for $c_1 = 0$.

%%%%%%%%%%%%%%%%%%%%%%%%%%%%%%%%%%%%%%
\section{Conclusions}
We have studied a possibility that one of the MSSM flat directions plays the role of the inflaton and
realizes the large-field  inflation, indicated by the recent detection of the primordial 
$B$-mode polarization by BICEP2. To this end we have imposed a generalized shift symmetry
on the flat direction, which enables us to control the inflaton potential over super-Planckian
field values. The flat direction inflation is realized in the context of the running kinetic inflation, 
where the kinetic term gets larger as  the field amplitude increases. It renders the potential effectively 
flatter to be consistent with the CMB observations. We have found that the inflaton could be 
the flat directions such as $LLe$, $LLddd$, or $QuQue$, which are lifted by the 
nonrenormalizable superpotential  of the form $W=X\phi^m$. A complete list of the possible flat 
directions as well as the predicted values of  $n_s$ and $r$ is given in the Appendix. 

The high-scale inflation, in general, predicts large baryonic/CDM isocurvature fluctuations for
those mechanisms of baryogenesis and/or dark matter creation by light scalar fields.
Our model, however, does not generate any baryonic isocurvature fluctuations, since the orthogonal
direction to the inflationary trajectory becomes heavy during inflation. 

The helical motion of the flat direction after inflation automatically takes place due to the same
$\phi$-number breaking terms in the K\"ahler potential, which are allowed by the generalized shift symmetry. 
When the soft SUSY breaking mass term dominates the potential, the field naturally transforms into $Q$ balls. 
Thus the reheating proceeds through the $Q$-ball decay, and the reheating temperature would be rather low. 
Importantly, it is straightforward to reheat the standard model particles since the inflaton is one of the MSSM flat 
directions. We have found that a sufficient amount of the baryon asymmetry can be created. 
For unsuppressed $\phi$-number violating operators, the resultant baryon number of the universe 
is much larger than the observed value, but it would be a remedy of baryogenesis for those scenarios 
that need late-time entropy production. Alternatively, the baryon asymmetry can be suppressed for a 
sufficiently small $|c_1|$. 

%%%%%%%%%%%%%%%%%%%%%%%%%%%%%%%%%%%%%%%%%%
\section*{Acknowledgments}
The work is supported by JSPS Grant-in-Aid for Young Scientists (B) 
(No.23740206 [SK] and No.24740135 [FT]),
the Grant-in-Aid for Scientific Research on Innovative Areas (No.23104008 [FT]),  
Inoue Foundation for Science [FT], and also by
World Premier International Center Initiative (WPI Program), MEXT, Japan.

\appendix

\section{Flat directions for the running kinetic inflation}
Here we show the complete list of the flat directions which can be the candidate for the inflaton in our model.
In Table~\ref{flat_table2}, we display the gauge-invariant (GI) monomials $(\phi^{n'})$ representing the 
corresponding flat directions, the nonrenormalizable superpotentials of the form of $X\phi^m$, 
which lift all the degrees of freedom in that direction, and the powers of the field $\hat{\phi} (= \phi^n)$ of
the effective potential $V \sim \hat{\phi}^p$ in the cases for $n=n'$ and $n=2n'$. 
For both cases, the predicted values of the scalar spectral index $n_s$ and 
the tensor-to-scalar ratio $r$ are shown in Table~\ref{lamnsr}, together with $\lambda$ in 
Eq.(\ref{superpot}), which is necessary for the Planck normalization on the curvature perturbations. 
See Eqs.(\ref{lambda}) - (\ref{t2s}).

\begin{table}[h!]
\caption{Flat direction candidates. $n=n'$ and $2n'$ cases are shown.}
\begin{tabular}{|c|c|c|c|c|}
\hline
Direction & GI monomials ($\phi^{n'}$) & $W_{\rm NR} (=X\phi^m) $ & $p=2m/n'$ & $p=2m/2n'$ \\
\hline
$L$, $d$ & $LLddd$ ($n'=5$) & $H_u LLLddd$ ($m=6$) & 12/5 = 2.4  & 12/10 = 1.2 \\
\hline
$Q$, $u$ & $QQQQu$ ($n'=5$) 
& $\left\{\begin{array}{l} QQQL \ (m=3) \\ QuQd \ (m=3) \end{array} \right.$  
& 6/5 = 1.2 & 6/10 = 0.6 \\
\hline
$Q$, $u$, $e$ & $QuQue$ ($n'=5$) & $H_dQuQuQuee$ ($m=8$) & 16/5=3.2 & 16/10=1.6 \\
\hline
$L$, $u$, $d$ & $\left\{\begin{array}{l} udd \ (n'=3) \\ LLddd \ (n'=5) \end{array} \right.$ 
& $\left\{\begin{array}{l} uddQdL \ (m=5) \\ LLeudd \ (m=5) \end{array} \right.$  
& $\begin{array}{l} 10/3=3.33 \\ 10/5=2 \end{array}$ & $\begin{array}{l} 10/6=1.67 \\ 10/10=1 \end{array}$ \\
\hline
$L$, $d$, $e$ & $\left\{\begin{array}{l} LLe \ (n'=3) \\ LLddd \ (n'=5) \end{array} \right.$ 
& $\left\{\begin{array}{l} H_d Lddd \ (m=4) \\ H_uLLLe \ (m=4) \end{array} \right.$  
& $\begin{array}{l} 8/3=2.67 \\ 8/5=1.6 \end{array}$ & $\begin{array}{l} 8/6=1.33 \\ 8/10=0.8 \end{array}$ \\
\hline
$L$, $u$, $e$ & $\left\{\begin{array}{l} LLe \ (n'=3) \\ uuuee \ (n'=5) \end{array} \right.$ 
& $\left\{\begin{array}{l} QuLe \ (m=3) \\ uude \ (m=3) \end{array} \right.$  
& $\begin{array}{l} 6/3=2 \\ 6/5=1.2 \end{array}$ & $\begin{array}{l} 6/6=1 \\ 6/10=0.6 \end{array}$ \\
\hline
$Q$, $L$, $e$ & $\left\{\begin{array}{l} LLe \ (n'=3) \\ QQQL \ (n'=4) \\ (QQQ)_4LLLe \ (n'=7) \end{array} \right.$
& $QuLe$ ($m=3$) & $\begin{array}{l} 6/3=2 \\ 6/4=1.5 \\ 6/7= 0.86 \end{array}$ 
& $\begin{array}{l} 6/6=1 \\ 6/8=0.75 \\ 6/14= 0.43 \end{array}$\\
\hline
$Q$, $L$, $d$ & $\left\{\begin{array}{l} QLd \ (n'=3) \\ QQQL \ (n'=4) \\ dddLL \ (n'=5) \end{array} \right.$
& $QuQd$ ($m=3$) & $\begin{array}{l} 6/3=2 \\ 6/4=1.5 \\ 6/5= 1.2 \end{array}$ 
& $\begin{array}{l} 6/6=1 \\ 6/8=0.75 \\ 6/10= 0.6 \end{array}$ \\
\hline
$Q$, $L$, $u$ & $\left\{\begin{array}{l} QQQL \ (n'=4) \\ QQQQu \ (n'=5) \end{array} \right.$ 
& $\left\{\begin{array}{l} QuLe \ (m=3) \\ QuQd \ (m=3) \end{array} \right.$  
& $\begin{array}{l} 6/4=1.5 \\ 6/5=1.2 \end{array}$ & $\begin{array}{l} 6/8=0.75 \\ 6/10=0.6 \end{array}$ \\
\hline
$Q$, $u$, $d$ & $\left\{\begin{array}{l} udd \ (n'=3) \\ QuQd \ (n'=4) \\ QQQQu \ (n'=5) \\ uudQdQd \ (n'=7) \end{array} \right.$
& $\left\{\begin{array}{l} QQQL \ (m=3) \\ uude \ (m=3) \end{array} \right.$  
& $\begin{array}{l} 6/3=2 \\ 6/4=1.5 \\ 6/5= 1.2 \\ 6/7=0.86 \end{array}$ 
& $\begin{array}{l} 6/6=1 \\ 6/8=0.75 \\ 6/10= 0.6 \\ 6/14=0.43 \end{array}$ \\
\hline
$L$, $u$, $d$, $e$ & $\left\{\begin{array}{l} udd \ (n'=3) \\ LLe \ (n'=3) \\ uude \ (n'=4) \\ 
dddLL \ (n'=5) \\ uueeu \ (n'=5)  \end{array} \right.$
& $Qude \ (m=3)$  
& $\begin{array}{l} 6/3=2 \\ 6/4=1.5 \\ 6/5= 1.2 \end{array}$
& $\begin{array}{l} 6/6=1 \\ 6/8=0.75 \\ 6/10= 0.6 \end{array}$ \\
\hline
$Q$, $L$, $d$, $e$ & $\left\{\begin{array}{l} LLe \ (n'=3) \\ QdL \ (n'=3) \\ QQQL \ (n'=4) \\ 
dddLL \ (n'=5) \\ (QQQ)_4LLLe \ (n'=7) \end{array} \right.$ 
& $\left\{\begin{array}{l} QuLe \ (m=3) \\ QuQd \ (m=3) \end{array} \right.$  
& $\begin{array}{l} 6/3=2 \\ 6/4=1.5 \\ 6/5=1.2 \\ 6/7=0.86 \end{array}$
& $\begin{array}{l} 6/6=1 \\ 6/4=0.75 \\ 6/10=0.6 \\ 6/14=0.43 \end{array}$ \\
\hline
$Q$, $L$, $u$, $e$ & $\left\{\begin{array}{l} LLe \ (n'=3) \\ QQQL \ (n'=4) \\ 
QuLe \ (n'=4) \\ uuuee \ (n'=5) \\ QuQue \ (n'=5) \\ QQQQu \ (n'=5) \\ (QQQ)_4LLLe \ (n'=7) \end{array} \right.$ 
& $\left\{\begin{array}{l} QuLe \ (m=3) \\ uude \ (m=3) \end{array} \right.$  
& $\begin{array}{l} 6/3=2 \\ 6/4=1.5 \\ 6/5=1.2 \\ 6/7=0.86 \end{array}$
& $\begin{array}{l} 6/6=1 \\ 6/8=0.75 \\ 6/10=0.6 \\ 6/14=0.43 \end{array}$ \\
\hline
$Q$, $u$, $d$, $e$ & $\left\{\begin{array}{l} udd \ (n'=3) \\ QuQd \ (n'=4) \\ 
uude \ (n'=4) \\ uuuee \ (n'=5) \\ QuQue \ (n'=5) \\ QQQQu \ (n'=5) \\ uudQdQd \ (n'=7) \end{array} \right.$ 
& $\left\{\begin{array}{l} QuLe \ (m=3) \\ QQQL \ (m=3) \end{array} \right.$  
& $\begin{array}{l} 6/3=2 \\ 6/4=1.5 \\ 6/5=1.2 \\ 6/7=0.86 \end{array}$
& $\begin{array}{l} 6/6=1 \\ 6/4=0.75 \\ 6/10=0.6 \\ 6/14=0.43 \end{array}$ \\
\hline
$Q$, $L$, $u$, $d$ & $\left\{\begin{array}{l} udd \ (n'=3) \\ QdL \ (n'=3) \\ QuQd \ (n'=4) \\ 
QQQL \ (n'=4) \\ dddLL \ (n'=5) \\ QQQQu \ (n'=5) \\ uudQdQd \ (n'=7) \end{array} \right.$ 
& $\left\{\begin{array}{l} QuLe \ (m=3) \\ uude \ (m=3) \end{array} \right.$  
& $\begin{array}{l} 6/3=2 \\ 6/4=1.5 \\ 6/5=1.2 \\ 6/7=0.86 \end{array}$
& $\begin{array}{l} 6/6=1 \\ 6/8=0.75 \\ 6/10=0.6 \\ 6/14=0.43 \end{array}$ \\
\hline
\end{tabular}
\label{flat_table2}
\end{table}

\begin{table}[htdp]
\caption{Scalar spectral index $n_s$, tensor-to-scalar ratio $r$, and the coupling $\lambda$.}
\begin{center}
\begin{tabular}{ccc}
\begin{tabular}{|c|c|c|c|c|c|}
\hline
$\ m \ $ & $\ n=n' \ $ & $\quad N \quad $ &  \hspace{8mm} $\lambda$ \hspace{8mm} 
& $\quad n_s \quad $ & $\quad r \quad$  \\
\hline
& & 40 & $1.16\times 10^{-5}$ & 0.950 & 0.200 \\
3 & 3 & 50 & $9.26\times 10^{-6}$ & 0.960 & 0.160 \\
& & 60 & $7.72\times 10^{-6}$ & 0.967 & 0.133 \\
\hline
& & 40 & $1.93\times 10^{-5}$ & 0.956 & 0.150 \\
3 & 4 & 50 & $1.59\times 10^{-5}$ & 0.965 & 0.120 \\
& & 60 & $1.35\times 10^{-5}$ & 0.971 & 0.100 \\
\hline
& & 40 & $2.51\times 10^{-5}$ & 0.960 & 0.120 \\
3 & 5 & 50 & $2.10\times 10^{-5}$ & 0.968 & 0.096 \\ 
& & 60 & $1.82\times 10^{-5}$ & 0.973 & 0.080 \\
\hline
& & 40 & $3.18\times 10^{-5}$ & 0.964 & 0.086 \\
3 & 7 & 50 & $2.71\times 10^{-5}$ & 0.9771 & 0.069 \\
& & 60 & $2.38\times 10^{-5}$ & 0.976 & 0.057 \\
\hline
& & 40 & $5.32\times 10^{-6}$ & 0.942 & 0.267 \\
4 & 3 & 50 & $4.10\times 10^{-6}$ & 0.953 & 0.213 \\
& & 60 & $3.31\times 10^{-6}$ & 0.961 & 0.178 \\
\hline
& & 40 & $1.75\times 10^{-5}$ & 0.955 & 0.160 \\
4 & 5 & 50 & $1.44\times 10^{-5}$ & 0.964 & 0.128 \\
& & 60 & $1.22\times 10^{-5}$ & 0.970 & 0.107 \\
\hline
& & 40 & $2.27\times 10^{-6}$ & 0.933 & 0.333 \\
5 & 3 & 50 & $1.68\times 10^{-6}$ & 0.947 & 0.267 \\
& & 60 & $1.32\times 10^{-6}$ & 0.956 & 0.222 \\
\hline
& & 40 & $1.16\times 10^{-5}$ & 0.950 & 0.200 \\
5 & 5 & 50 & $9.26\times 10^{-6}$ & 0.960 & 0.160 \\
& & 60 & $7.72\times 10^{-6}$ & 0.967 & 0.133 \\
\hline
& & 40 & $7.33\times 10^{-6}$ & 0.945 & 0.240 \\
6 & 5 & 50 & $5.74\times 10^{-6}$ & 0.956 & 0.192 \\
& & 60 & $4.70\times 10^{-6}$ & 0.963 & 0.160 \\
\hline
& & 40 & $2.70\times 10^{-6}$ & 0.935 & 0.320 \\
8 & 5 & 50 & $2.02\times 10^{-6}$ & 0.948 & 0.256 \\
& & 60 & $1.59\times 10^{-6}$ & 0.957 & 0.213 \\
\hline
\end{tabular}
& \hspace{5mm} &
\begin{tabular}{|c|c|c|c|c|c|}
\hline
$\ m \ $ & $\ n=2n' \ $ & $\quad N \quad $ &  \hspace{8mm} $\lambda$ \hspace{8mm} 
& $\quad n_s \quad $ & $\quad r \quad$  \\
\hline
& & 40 & $2.91\times 10^{-5}$ & 0.963 & 0.100 \\
3 & 6 & 50 & $2.46\times 10^{-5}$ & 0.970 & 0.080 \\
& & 60 & $2.15\times 10^{-5}$ & 0.975 & 0.067 \\
\hline
& & 40 & $3.35\times 10^{-5}$ & 0.966 & 0.075 \\
3 & 8 & 50 & $2.87\times 10^{-5}$ & 0.973 & 0.060 \\
& & 60 & $2.54\times 10^{-5}$ & 0.977 & 0.050 \\
\hline
& & 40 & $3.52\times 10^{-5}$ & 0.968 & 0.060 \\
3 & 10 & 50 & $3.05\times 10^{-5}$ & 0.974 & 0.048 \\ 
& & 60 & $2.71\times 10^{-5}$ & 0.978 & 0.040 \\
\hline
& & 40 & $3.53\times 10^{-5}$ & 0.970 & 0.043 \\
3 & 14 & 50 & $3.09\times 10^{-5}$ & 0.9776 & 0.034 \\
& & 60 & $2.76\times 10^{-5}$ & 0.980 & 0.029 \\
\hline
& & 40 & $2.25\times 10^{-5}$ & 0.958 & 0.133 \\
4 & 6 & 50 & $1.86\times 10^{-5}$ & 0.967 & 0.107 \\
& & 60 & $1.60\times 10^{-5}$ & 0.961 & 0.178 \\
\hline
& & 40 & $3.27\times 10^{-5}$ & 0.965 & 0.080 \\
4 & 10 & 50 & $2.80\times 10^{-5}$ & 0.972 & 0.064 \\
& & 60 & $2.47\times 10^{-5}$ & 0.977 & 0.053 \\
\hline
& & 40 & $1.64\times 10^{-5}$ & 0.954 & 0.167 \\
5 & 6 & 50 & $1.34\times 10^{-5}$ & 0.963 & 0.133 \\
& & 60 & $1.13\times 10^{-6}$ & 0.970 & 0.111 \\
\hline
& & 40 & $2.91\times 10^{-5}$ & 0.963 & 0.100 \\
5 & 10 & 50 & $2.46\times 10^{-5}$ & 0.970 & 0.080 \\
& & 60 & $2.15\times 10^{-5}$ & 0.975 & 0.067 \\
\hline
& & 40 & $2.51\times 10^{-5}$ & 0.960 & 0.120 \\
6 & 10 & 50 & $2.10\times 10^{-5}$ & 0.968 & 0.096 \\
& & 60 & $1.82\times 10^{-5}$ & 0.973 & 0.080 \\
\hline
& & 40 & $1.75\times 10^{-5}$ & 0.955 & 0.160 \\
8 & 10 & 50 & $1.44\times 10^{-5}$ & 0.964 & 0.128 \\
& & 60 & $1.22\times 10^{-5}$ & 0.970 & 0.107 \\
\hline
\end{tabular}
\end{tabular}
\end{center}
\label{lamnsr}
\end{table}

%%%%%%%%%%%%%%%%%%%%%%%%%%%%%%%%%%%%

%%%%%%%%%%%%%%%%%%%%%%%%%%%%%%%%%%%%

\end{document}